\documentclass[prl,aps,twocolumn,showpacs,floats,amssymb,amsfonts]{revtex4}
\usepackage{graphicx}
\usepackage{bm}

\begin{document}

\title{
The Density of States in High-$\bm{T_c}$ Superconductors Vortices
}

\author{
C. Berthod and B. Giovannini
}

\affiliation{
DPMC, Universit\'e de Gen\`eve, 24 Quai Ernest-Ansermet,
1211 Gen\`eve 4, Switzerland
}

\date{May 22, 2001}

\begin{abstract}

We calculated the electronic structure of a vortex in a pseudogapped
superconductor within a model featuring strong correlations. With increasing
strength of the correlations, the BCS core states are suppressed and the
spectra in and outside the core become similar. If the correlations are
short-range, we find new core states in agreement with the observations in
YBa$_2$Cu$_3$O$_{7-\delta}$ and Bi$_2$Sr$_2$CaCu$_2$O$_{8+\delta}$. Our results
point to a common phenomenology for these two systems and indicate that
normal-state correlations survive below $T_c$ without taking part in the
overall phase coherence.

\end{abstract}

\maketitle

Among the anomalous properties of high-temperature superconductors (HTS), one
of the most intriguing is the pseudogap phenomenon observed in the normal state
of underdoped systems. The pseudogap state is marked by a substantial decrease
of the one-particle density of states near the Fermi energy, and can be
observed in several spectroscopic, thermodynamic, and transport experiments
(see Ref.~\onlinecite{Timusk-99} for a review). Its elucidation is widely
thought of as one of the keys to the understanding of high-temperature
superconductivity. Unfortunately, detailed studies of the low-lying excitations
in the pseudogap state are impossible due to the emergence of the
superconducting state at low temperature. Vortex-core spectroscopy is perhaps
one way into this problem: it has indeed been stated, on the basis of
experimental findings\cite{Renner-98}, that the local density of states (LDOS)
in vortex cores below $T_c$ reflects the properties of the (pseudogapped)
normal state above $T_c$. More precisely one can argue that the physical state
inside vortices should reflect the properties of the normal state at low
temperature in the boundary conditions set by a gradual vanishing of the
superconducting order in the core. A detailed analysis of this density of
states may therefore lead to useful insights into the nature of the pseudogap
state itself.

Some characteristics of the vortex cores seem to be now well established in
YBa$_2$Cu$_3$O$_{7-\delta}$ (YBCO) and Bi$_2$Sr$_2$CaCu$_2$O$_{8+\delta}$
(BSCCO), the two systems for which experimental results are
available\cite{Renner-98,Aprile-95,Hoogenboom-00,Pan-00}: the LDOS in the core
does not display a central peak as predicted by the BCS theory, but
particle-hole symmetric low-energy states; noticeably, the amplitude of the
core states is much larger in YBCO than in BSCCO. In BSCCO, moreover, the core
LDOS shows remarkable similarity (when properly thermally broadened) to the
density of states observed above $T_c$ in the pseudogap regime.

Theoretically, a number of calculations have been done in the past years and
competing models of HTS lead to increasingly detailed predictions. The first
calculations were undertaken within the Bogoliubov--de Gennes formulation of
the BCS theory\cite{Soininen-94,Wang-95,Maki-97,Franz-98,Yasui-99}. For a
$d_{x^2-y^2}$ vortex the LDOS shows a broad zero-bias conductance peak (ZBCP)
and a four-fold symmetric shape. This picture is not consistent with the
experimental observations in YBCO and BSCCO. In another BCS-type scenario, a
secondary order parameter component of a different symmetry is generated
locally by the magnetic field and the resultant pair potential no longer has
nodes: the central peak would thus be suppressed or split\cite{Franz-98}. The
absence of ZBCP has also been tentatively explained by the properties of the
$c$-axis tunneling matrix elements\cite{Wu-00}. Many workers in the field
believe however that the absence of central peak cannot be explained in the BCS
framework, and needs some kind of fundamental extension of the BCS theory that
would take into account strong correlations. The first attempts were made
within the SO(5) theory and predicted an insulating
core\cite{Arovas-97,Andersen-00}, but the detailed predictions of this model
seem not to agree with experiment. Some $t$-$J$ model calculations have
predicted a ZBCP split by an induced $s$-wave order parameter at low
doping\cite{Himeda-97,Han-00}. Very recently, calculations based on the SU(2)
slave-boson theory were reported\cite{Kishine-01}. The authors make detailed
predictions for the LDOS in vortex cores, which do not seem to correspond to
the experimental data available today and which should allow to test in a
significant way the slave-boson theory of HTS.

In this Letter we present calculations based on the Cooperon-propagator
description of HTS. We show that this approach can explain several
characteristics of the vortex-core spectra in YBCO and BSCCO. In addition, it
allows us to give a tentative explanation for one of the (many) puzzles of HTS,
namely whether the gap and pseudogap are of similar (superconducting) origin or
whether they correspond to completely different physical processes. This point
will be commented upon in the discussion below. First, we briefly review the
Cooperon-propagator description of superconductivity and the model used in the
numerical calculations. In the Kadanoff-Martin theory of
superconductivity\cite{Kadanoff-61}, which is entirely based on correlation
functions, the electronic self-energy takes the
form\cite{Giovannini-01}\pagebreak
	\begin{eqnarray}\label{eq:Sigma}
		\nonumber \Sigma(\bm{r},\bm{s},\tau) &=&
		-\int\!d\bar{\bm{r}}d\bar{\bm{s}}\,
		\langle T_{\tau}\{\Delta(\bm{r},\bar{\bm{r}},\tau)
		\Delta^{\dagger}(\bar{\bm{s}},\bm{s},0)\} \rangle\\
		&&\times\,{\cal G}_0(\bar{\bm{s}}\!-\!\bar{\bm{r}},-\tau),
	\end{eqnarray}
where $\Delta(\bm{r},\bm{s},\tau) = V(\bm{r}\!-\!\bm{s})
\psi_{\uparrow}(\bm{r},\tau) \psi_{\downarrow}(\bm{s},\tau)$, $V(\bm{r})$ is
the pairing interaction, and ${\cal G}_0(\bm{r},\tau)$ is the free propagator.
In this formalism the superconducting order is characterized by the long-range
properties of the pair correlation function. Eq.~(\ref{eq:Sigma}) is valid both
above and below $T_c$ provided higher order correlation effects are not
important, and we will assume that it remains valid regardless of the model or
approximations involved in calculating $\langle\Delta\Delta^{\dagger}\rangle$.
In particular, the BCS theory is recovered by setting
$\langle\Delta\Delta^{\dagger}\rangle =
\langle\Delta\rangle\langle\Delta^{\dagger}\rangle$ where
$\langle\Delta\rangle$ is the BCS order parameter. Following recent
studies\cite{Giovannini-01} we model the correlation function (Cooperon
propagator) through:
	\begin{eqnarray}\label{eq:model}
		\nonumber
		\langle T_{\tau}\{\Delta\Delta^{\dagger}\}\rangle &=&
		\Delta_p^{ }(\bm{r}\!-\!\bar{\bm{r}})
		R(\ell/\varrho)
		\Delta_p^{ }(\bar{\bm{s}}\!-\!\bm{s})\\
		&&+\left\{\begin{array}{ll}
		\!\!\!\Delta_s^{ }(\bm{r}\!-\!\bar{\bm{r}})
		F(\ell/\xi)
		\Delta_s^{ }(\bar{\bm{s}}\!-\!\bm{s})&T>T_c\\
		\!\!\!\Delta_s^{ }(\bm{r},\bar{\bm{r}})
		\Delta_s^{\star}(\bar{\bm{s}},\bm{s})&T\leqslant T_c.
		\end{array}\right.
	\end{eqnarray}
The first term characterizes the strength and symmetry ($\Delta_p$) and the
range ($\varrho$) of the normal-state correlations leading to the pseudogap
below the temperature $T^*$, with $\ell=\frac{1}{2}
|\bm{r}\!+\!\bar{\bm{r}}\!-\!\bar{\bm{s}}\!-\!\bm{s}|$ and $R(x)=e^{-x}$. The
second term accounts for the phase-coherence properties of the superconductor
and the phase fluctuations responsible for destroying superconductivity. Above
$T_c$, $\Delta_s$ must be understood as the modulus of a local BCS-like order
parameter. The phase physics is contained in the function $F(x)\sim e^{-x}$
which is the phase-phase correlation function. $F$ thus reduces to a function
of the distance $\ell$ with a correlation length $\xi(T)$ diverging at $T_c$ in
the Kosterlitz-Thouless fashion. Below $T_c$, $F$ is assumed to factorize, as
in the BCS theory, into the product of two local phases. These phases are
embodied in the amplitude $\Delta_s$ which becomes position dependent and
complex. A more elaborate model should include dynamic correlations. We showed,
however, that Eq.~(\ref{eq:model}) leads to reasonable agreement with some
aspects of the experimental density of states as a function of $T$ in
homogeneous systems\cite{Giovannini-01}. Eq.~(\ref{eq:model}) may be viewed as
a non-local generalization of the Ansatz $\Delta^2=\Delta_p^2+\Delta_s^2$
introduced in Ref.~\onlinecite{Chen-98}.

In what follows we concentrate on the vortex state at $T=0$, then $\Delta_s$
plays the role of a BCS gap and vanishes in the vortex cores while its phase
winds by $2\pi$ in response to the applied field. Eq.~(\ref{eq:model})
basically means that the Cooperon propagator has a short-range part in addition
to the long-range part described by $\Delta_s$. Our central assumption is that
these short-range normal-state correlations are not involved in the phase
coherence and remain finite in the cores. This view is supported by recent NMR
experiments showing that the pseudogap is not affected by a strong magnetic
field\cite{Zheng-00}. We also note that similar considerations apply to the
spinon and holon fields in theories based on spin-charge
separation\cite{Lee-00,Franz-01}.

\begin{figure}[b!]
\includegraphics[width=8.6cm]{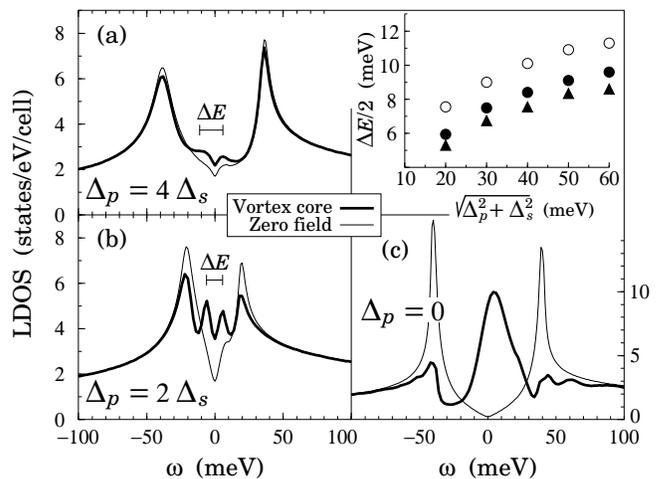}
\caption{\label{fig:main}
Vortex-core and zero-field DOS for the parameters $\varrho=5a$ and (a)
$\Delta\equiv\sqrt{\Delta_p^2+\Delta_s^2}=40$~meV, $\Delta_p=4\Delta_s$; (b)
$\Delta=20$~meV, $\Delta_p=2\Delta_s$. (c) BCS limit: $\Delta_p=0$ and
$\Delta_s=40$~meV. Inset: Halved core-level separation as a function of
$\Delta$ for $\varrho=5a$ (filled symbols) and $\varrho=10a$ (empty symbols).
Circles are for $\Delta_p=2\Delta_s$ and triangles for $\Delta_p=4\Delta_s$. No
core states are found for $\varrho=10a$ and $\Delta_p=4\Delta_s$.
}
\end{figure}

Using Eqs.~(\ref{eq:Sigma}) and (\ref{eq:model}) we calculate the Green's
function for an isolated $d_{x^2-y^2}$ vortex on a two-dimensional lattice. We
neglect Landau-level quantization, as appropriate for low magnetic fields. The
modulus of $\Delta_s$ is proportional to $\tanh(r/a)$, with $a$ the lattice
parameter. The pseudogap $\Delta_p$ has $d_{x^2-y^2}$ symmetry in agreement
with photoemission data\cite{Ding-96} and is assumed uniform in space for
simplicity. The self-energy is rewritten as $\Sigma(\omega)=\Sigma_{\rm
u}(\omega)+\Sigma_{\rm i}(\omega)$ where $\Sigma_{\rm u}$ is uniform and
$\Sigma_{\rm i}$ contains the inhomogeneous term. The uniform part of the
Green's function is evaluated as ${\cal G}_{\rm u}=({\cal G}_0^{-1}-\Sigma_{\rm
u})^{-1}$ in reciprocal space, using a $1024\times 1024$ $k$-point mesh. The
band structure is taken from Ref.~\onlinecite{Norman-95} with the Fermi level
$E_{\rm F}$ located 10~meV above the van-Hove singularity. This yields
zero-field DOS's in qualitative agreement with experiment. The Green's function
is finally evaluated as ${\cal G}=({\cal G}_{\rm u}^{-1}-\Sigma_{\rm i})^{-1}$
on a $M\times M$ real-space mesh with the vortex at the center. This step
necessitates the inversion of two large matrices ($M^2\times M^2$) for every
energy $\omega$. We have checked that the calculations are well converged with
$M=41$ and reproduce the published results of the BCS theory when $\Delta_p=0$
[see Fig.~\ref{fig:main}(c)].

Our main results are presented in Fig.~\ref{fig:main}. Fig.~\ref{fig:main}(a)
shows the vortex-core LDOS $N(\bm{r}=0,\omega)$ and the zero-field DOS
$N_0(\omega)$ for a pseudogapped superconductor. Instead of a ZBCP, we find two
core states in the vortex spectrum. Our numerical analysis shows that the
subgap peaks appear only for moderate values of $\Delta_p/\Delta_s$ and low
values of $\varrho$. These peaks grow as $\Delta_p/\Delta_s$ and/or $\varrho$
decrease, and eventually merge to form the broad BCS peak as
$\Delta_p\rightarrow 0$. An intermediate example is displayed in
Fig.~\ref{fig:main}(b). In fact, weak subgap structures exist also in the
zero-field DOS: the vortex seems to induce a localization of preexisting
low-energy states. We therefore expect those pseudogap-induced core states to
be independent of magnetic field as observed in recent
experiments\cite{Hoogenboom-01}. On the contrary the BCS core states were shown
to split at high vortex density\cite{Yasui-99}; a field-dependent splitting is
also expected in scenarios based on a secondary $id_{xy}$ order
parameter\cite{Franz-98}.

It was found experimentally that the coherence peaks are suppressed in the
core\cite{Renner-98,Aprile-95,Hoogenboom-00,Pan-00}, unlike in our results,
Fig.~\ref{fig:main}(a) and (b). In our calculations the spectral weight in a
small energy window ($\sim 4\Delta$) around $E_{\rm F}$ is locally conserved,
in contrast to experiments. This discrepancy might be due to a vortex charging
effect or to our assumption of static correlations in the model
Eq.~(\ref{eq:model}). Although, at this stage, the model is unable to account
for the observed behavior of the coherence peaks in the core, we tentatively
argue below that it can well explain the properties of the core states in both
YBCO and BSCCO.

The energy $\Delta E$ of the core states is plotted in the inset of
Fig.~\ref{fig:main} as a function of the gap scale
$(\Delta_p^2+\Delta_s^2)^{\frac{1}{2}}$. The calculated core-state energies
increase with the gap width and the correlation length $\varrho$ but depend
little on $\Delta_p/\Delta_s$. They are in good general agreement with the
experimental observations, i.e., 5.5~meV in a YBCO sample with a gap of
20~meV\cite{Aprile-95}, 7~meV for a 32~meV-gap BSCCO\cite{Pan-00}, and 14~meV
for a 45~meV-gap BSCCO\cite{Hoogenboom-00}. A recent thorough investigation of
the BSCCO system indicates a roughly linear relationship between $\Delta E$ and
the energy gap\cite{Hoogenboom-01}. Moreover, the fact that the observed peaks
are sharper in YBCO than in BSCCO can be well explained by assuming a smaller
value of $\Delta_p/\Delta_s$ in YBCO [see Fig.~\ref{fig:main}(b)]. This
assumption is consistent with the observation that optimally-doped YBCO, unlike
optimally-doped BSCCO, seems to have no pseudogap phase above
$T_c$\cite{Aprile-00}.

\begin{figure}[tb!]
\includegraphics[width=8.6cm]{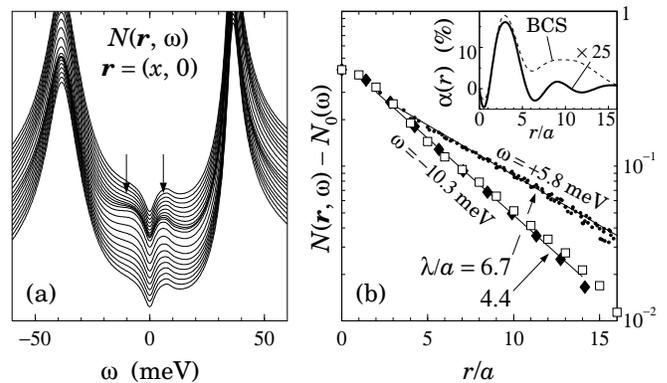}
\caption{\label{fig:spatial}
LDOS near the vortex core for the model parameters as in
Fig.~\ref{fig:main}(a). (a) $N(\bm{r},\omega)$ at all sites between $(-10,0)$
and $(10,0)$ (top to bottom). The curves are shifted vertically by equal
amounts. (b) Angular average of the LDOS (top) and LDOS along nodal
($\blacklozenge$) and anti-nodal ($\square$) directions at the energies
indicated by the arrows in (a). The strait lines show exponential behavior with
a decay length $\lambda$ as indicated. Inset: LDOS anisotropy ratio (see text);
smooth curves were obtained by interpolation of the lattice results.
}
\end{figure}

We now discuss the behavior of the LDOS in the vicinity of the core.
$N(\bm{r},\omega)$ is displayed in Fig.~\ref{fig:spatial}(a) as a function of
$\omega$ for $\bm{r}$ along $\hat{\bm{x}}$, illustrating the weak localization
of the core states. The angular average $\bar{N}(r,\,\omega)$ plotted in
Fig.~\ref{fig:spatial}(b) at the core-state energy shows the exponential decay
of the LDOS, as observed experimentally\cite{Pan-00} and in contrast to the
algebraic decay of BCS core states. The calculated decay length depends on
energy and is $\lambda/a=6.7$ for the state above $E_{\rm F}$ and 4.4 for the
state below $E_{\rm F}$. The average 5.6 compares well with the value of
$5.7\pm0.8$ quoted in Ref.~\onlinecite{Pan-00} (using $a=3.83$~\AA\@ for the
BSCCO lattice parameter). Fig.~\ref{fig:spatial}(b) also shows the absence of a
significant four-fold anisotropy in $N(\bm{r},\omega)$ around the core. It is
known from semiclassical approaches that a finite lifetime reduces the
anisotropy\cite{Franz-99}. The pseudogap correlations have the same effect in
our calculations. As an illustration we plot in the inset the ratio $\alpha(r)
= \int_0^{\Delta}d\omega\,[N(r\hat{\bm{n}},\omega) -
N(r\hat{\bm{x}},\omega)]/\int_0^{\Delta}d\omega\,\bar{N}(r,\omega)$ which
measures the relative anisotropy of the LDOS at subgap energies ($\hat{\bm{n}}$
is the nodal direction). In the BCS limit $\alpha(r)$ is maximum around
$r/a=3$, consistently with previous results\cite{Maki-97}. The behavior of
$\alpha(r)$ in the pseudogapped case is similar, although reduced by a factor
$\sim25$, making the maximum anisotropy smaller than 1\%.

Our model Cooperon propagator has deliberately few parameters. The good
agreement between numerical results and experimental data gives us confidence
that it may contain the relevant input for interpreting vortex-core
measurements in HTS. Our calculations show clearly that the same model and
parameters used in the pseudogap regime also explain the properties of the core
states in vortices. Whether the pseudogap part of the self-energy survives
unchanged in the superconducting state or whether it reappears gradually only
in the vortex cores should be considered an open problem, although our
calculations in the homogeneous system and in the vortex, as well as some
experimental evidence\cite{Renner-98,Krasnov-00}, seem to favor the first term
of the alternative. Moreover, the short-range part of the Cooperon propagator
only reproduces the experimental results if it is incoherent, i.e., not
participating in the overall phase coherence below $T_c$. This belongs today to
the conventional wisdom, essentially because this part does not vanish in the
center of vortices, and also because of the weak magnetic-field dependence of
the pseudogap\cite{Zheng-00}. Finally, our model involves two different length
scales, the superconducting coherence length and the range of the incoherent
correlations. We note here that recent experiments by Pan {\it et
al.\/}\cite{Pan-01} have shown the possible existence of two correlation
lengths in BSCCO, although in their interpretation the shortest one is
associated with the presence of the oxygen dopants in the BiO plane and is
therefore not translationally invariant as in our approach.

According to our picture the pseudogap and superconducting gap may have very
different properties, and this seems to be supported by
thermodynamic\cite{Loram-00} and reflectivity\cite{Basov-01} experiments. The
apparent contradiction with tunneling and photoemission data, which seem to
show that the superconducting gap merges smoothly with the pseudogap at $T_c$,
can be resolved if the one-electron energy gap below $T_c$ turns out to be
composed of both superconducting gap and pseudogap. This in fact happens in our
model, which interestingly produces a single gap structure in the DOS with a
smooth evolution across $T_c$, regardless of the relative values of $\Delta_p$
and $\Delta_s$\cite{Giovannini-01}.

Our theory being semi-phenomenological, it does not provide a deep
understanding of the underlying microscopic phenomena. It is compatible, in
principle, with any microscopic theory that would predict two energy scales ---
like the spin-charge separation scenario (which has the spinon gap and the
superconducting gap as separate entities) or the theory of stripe-induced
superconductivity (where, in one version, one-dimensional spinon field can
exist above $T_c$, the superconducting order being created by Josephson
coupling between the stripes) or the phase fluctuation picture, as emphasized
below --- and two length scales associated with these energy scales and
appearing in the calculation of the Cooperon propagator. The picture that
emerges for the pseudogap state in our model is analogous to a plane rotator
model, where the two-dimensional spins would be formed at some temperature
$T^*$, but where the phase coherence only concerns the interaction between
these spins (or preformed pairs), not the internal structure of the spins
themselves. In the model, however, the two-dimensional spins appear only in the
correlation function, not as entities located at definite sites. This picture,
when proper temperature dependence is included in $\Delta_s(T)$, could lead to
a second cross-over temperature within the pseudogap state, perhaps seen in
Nernst effect\cite{Xu-00}, conductivity\cite{Corson-99}, and Hall effect
measurements\cite{Matthey-01}, temperature at which local phase coherence is
established between neighboring spins, thus allowing the formation of
fluctuating vortices.

In conclusion, we display in this work a reasonably good agreement between a
phenomenological theory of the pseudogap state and the experimental data for
the vortex-core tunneling spectra in YBCO and BSCCO. Our results argue in favor
of a common, pseudogap related, mechanism at the origin of the core states in
these materials. The incoherence and short-rangeness of the pseudogap
correlations turn out to be the clue to the formation of these states, and
endow them with properties (energy, amplitude, spatial decay) which are
drastically different from the properties of conventional BCS core states.

We wish to thank \O. Fischer, B. W. Hoogenboom, S. H. Pan, and J.-M. Triscone
for valuable discussions.

\end{document}